\documentclass[doublecol]{epl2}

\usepackage{ifthen}
\usepackage{ifpdf}
\usepackage{color}

\ifpdf
\usepackage{graphicx}
\usepackage{epstopdf}
\else
\usepackage{graphicx}
\usepackage{epsfig}
\fi

\usepackage{latexsym}
\usepackage{amsmath}
\usepackage{amssymb}
\usepackage{bm}
\usepackage{wasysym}

\newcommand{\mass}{\mathsf{m}}

\newcommand{\tbox}[1]{\mbox{\tiny #1}}
 
\newcommand{\mylabel}[1]{\label{#1}} 
\newcommand{\be}[1]{\begin{eqnarray}\ifthenelse{#1=-1}{\nonumber}{\ifthenelse{#1=0}{}{\mylabel{e#1}}}}
\newcommand{\ee}{\end{eqnarray}}

\newcommand{\Eq}[1]{Eq.~(\ref{#1})} 
\newcommand{\Fig}[1]{\textcolor{blue}{Fig.\!\!~\ref{#1}}}  
\newcommand{\hide}[1]{} 
\newcommand{\rmrk}[1]{#1}
\newcommand{\mycite}[1]{\textcolor{blue}{\cite{#1}}}

\title{Quantum response of weakly chaotic systems}
\shorttitle{Driven chaotic systems}

\author{Alexander Stotland$^1$, Louis M. Pecora$^2$ and Doron Cohen$^1$}

\institute{
\mbox{$^1$Department of Physics, Ben-Gurion University, Beer-Sheva 84105, Israel}\\
\mbox{$^2$Code 6362, Naval Research Lab, Washington DC 20375, USA} 
}

\pacs{03.65.-w}{Quantum mechanics}

\abstract{
Chaotic systems, that have a small Lyapunov exponent, 
do not obey the common random matrix theory predictions 
within a wide ``weak quantum chaos'' regime.   
This leads to a novel prediction for the rate of heating   
for cold atoms in optical billiards with vibrating walls.
The Hamiltonian matrix of the driven system does not 
look like one from a Gaussian ensemble, but rather it is very sparse. 
This sparsity can be characterized by parameters~$s$ and~$g_s$ 
that reflect the percentage of large elements, and their connectivity 
respectively. For~$g_s$ we use a resistor network calculation that has 
direct relation to the semi-linear response characteristics of the system. 
}

\begin{document}
\maketitle


The heating of particles in a box with vibrating walls 
is a prototype problem for exploring the limitations 
of linear response theory (LRT) and the quantum-to-classical 
correspondence (QCC) principle. In the experimental arena 
this topic arises in the theory of {\em nuclear friction}~\mycite{wall1}, 
and more recently in the studies of cold atoms 
that are trapped in {\em optical billiards}~\mycite{nir1}.   
It is also related to the analysis
of mesoscopic conductance of ballistic rings~\mycite{bld}.   
Formally the dynamics is generated by a time 
dependent Hamiltonian $\mathcal{H}[f(t)]$, 
where $f(t)$ parametrizes the displacement of boundary, 
analogous to the time dependent electric field 
of the conductance problem.
In typical circumstances the classical analysis predicts 
an {\em absorption coefficient}~$G$ that is determined by 
the Kubo formula~\mycite{ott1,jar1,wilk,robbins,crs}, 
leading to the ``Wall formula" in the nuclear context, 
or to the analogous ``Drude formula" in the mesoscopic context.

\rmrk{If upon quantization we get for the absorption 
coefficient an $\hbar$~dependent result, 
that does not correspond to the classical result, 
we call it an {\em anomaly}.}
The question arises what are the circumstances 
in which anomalies show up~\mycite{wilk,robbins,crs,krav1,kbr,slr,kbw}.
There are {\em ``microscopic circumstances"} in which an anomaly 
is not a big surprise:
{\bf (1)} If $f(t)$ is slowly varying,   
so-called quantum adiabatic parametric driving, 
then Landau-Zener transitions between 
neighboring levels might be the \rmrk{dominant} 
mechanism for heating~\mycite{wilk}, 
and hence QCC is not expected. 
{\bf (2)} If $f(t)$ is low frequency noisy driving, 
that induces Fermi-Golden-Rule (FGR) transitions 
between neighboring levels only, 
the result would be determined by the level spacing statistics, 
and hence QCC is not expected~\mycite{slr}.

In this Letter we identify a ``weak quantum chaos regime" 
where a quantum anomaly shows up in quite typical {\em ``mesoscopic circumstances"},  
where QCC would be expected by common-wisdom.    

\vspace{2mm}

{\bf \rmrk{Modeling.--} } 
We consider a weakly chaotic billiard that has 
linear size~$L$ and a convex wall of radius~$R$. 
The Hamiltonian can be written schematically as 
\be{1}
\mathcal{H}[f(t)] \ = \ \mathcal{H} - f(t) F \ = \ \mathcal{H}_0 + U - f(t) F
\ee 
\rmrk{Specifically with regard to the numerical example of \Fig{fig:cqc:billiard}},  
$\mathcal{H}_0$ describes a non-deformed rectangular box
of length $L_x=L=1.5$ (upper edge), and width $L_y=1.0$.   
The term $U$~describes the deformation of the fixed (left) wall:
it is an arc of radius ${R=8}$ whose center of curvature 
is shifted upwards a vertical distance ${\Delta y =0.1}$ 
to break the reflection symmetry.    
The term $F$ is the perturbation due to the displacement~$f(t)$ 
of the moving (right) wall which can be regarded as a {\em piston}. 
Later we characterize the time dependence of~$f(t)$.  

Our interest is focused in circumstances 
in which the Lyapunov (correlation) time ${t_{\tbox{R}}=R/v_{\tbox{E}}}$ 
is much longer than the ballistic time ${t_{\tbox{L}}=L/v_{\tbox{E}}}$, 
where ${v_{\tbox{E}}=(2E/\mass)^{1/2}}$ is the velocity of the particle.
\rmrk{Turning to the quantum analysis we realize that the minimal 
model for~$\mathcal{H}$ depends on {\em two} dimensionless parameters:} 
\be{0}
u \ =&  L/R \ \ \ & \mbox{[dimensionless deformation]} \\
\hbar \ =& \lambda_{\tbox{E}}/L \ \ \ &\mbox{[dimensionless Planck const]}
\ee
Here $\lambda_{\tbox{E}}{=}2\pi\hbar_{\tbox{\rmrk{Planck}}}/(\mass v_{\tbox{E}})$ 
is the de~Broglie wavelength.  
For a given deformation ($R$~determines~$u$) 
and energy window ($E$~determines~$\hbar$)  
we calculate the eigenvalues and eigenfunctions of $\mathcal{H}$ 
using the boundary element method~\mycite{boundary}, 
find the ordered eigenenergies $E_n$, and calculate the matrix 
elements~$F_{nm}$ \rmrk{using the formula} 
\be{0}
F_{nm} = 
-\frac{1}{2\mass}\int \varphi^{(n)}(y)\varphi^{(m)}(y) \ dy
\ee
where $\varphi^{(n)}(y)$ is the normal derivative of 
the $n$th eigenfunction along the piston boundary.  
An image of a representative matrix is displayed in \Fig{fig:cqc:F:nm2:image}, 
and its bandprofile is presented in \Fig{fig:F:alg:vs:omega:deform}.

\begin{figure}
\includegraphics[clip,width=\hsize]{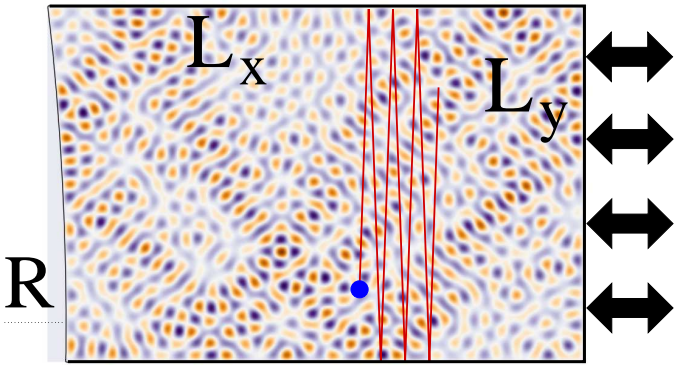}

\caption{
Sketch of the billiard system of \Eq{e1}. 
The unperturbed billiard is a \rmrk{rectangle} of size $L_x{=}1.5$ and $L_y{=}1.0$.  
The deformation $U$, due to the curvature of the left wall (radius ${R{=}8}$),  
is characterized by the parameter ${u=L_y/R}$. 
In order to break the mirror symmetry the center of the curved wall 
is shifted upwards a vertical distance ${\Delta y = 0.1}$.   
The time dependent perturbation is due to the displacement $f(t)$ of the right wall. 
In the numerics the units are chosen such that $\hbar_{\tbox{Planck}}{=}1$ 
and the mass is $\mass{=}1/2$. 
The image in the background represents  
the eigenstate $E_n{\simeq}13618$.
}

\mylabel{fig:cqc:billiard}
\end{figure}

\begin{figure}
\includegraphics[clip,width=\hsize]{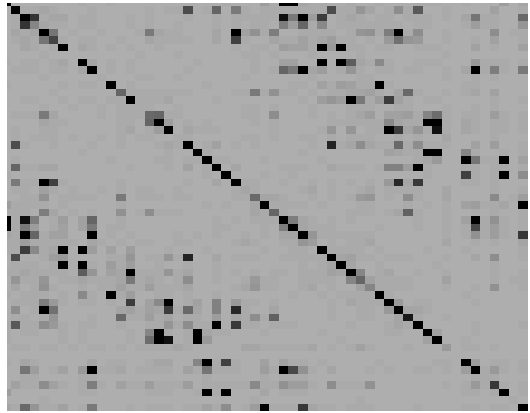}

\caption{
{\bf Image of the perturbation matrix.}
Image of the matrix $\bm{X}=\{|F_{nm}|^2\}$ for the billiard of Fig.~1
within the energy window ${3500< E_n < 4000}$. 
This matrix is {\em sparse}. More generally 
it might have some {\em texture}. The latter 
term applies if the arrangement of the large 
elements is characterized by some pattern. 
}

\mylabel{fig:cqc:F:nm2:image}
\end{figure}

\begin{figure}
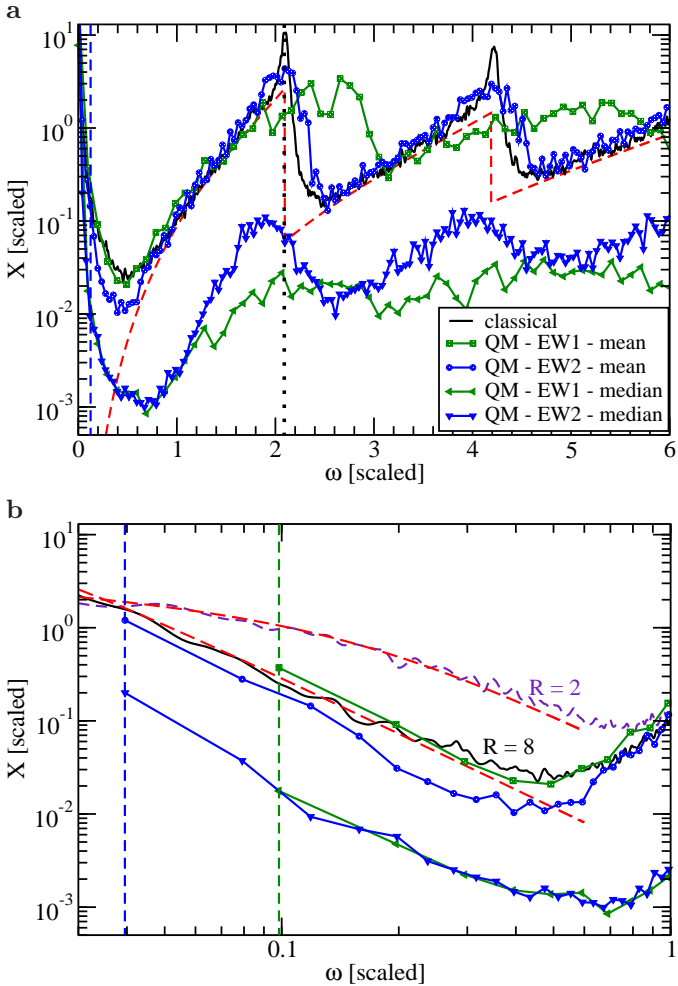

{\bf a} \hspace*{0.4\hsize} \\ 
\includegraphics[clip,width=\hsize]{cqc_F_alg_med_vs_omega_deform} \\
{\bf b}  \hspace*{0.4\hsize}  \\ 
\includegraphics[clip,width=\hsize]{cqc_F_alg_med_vs_omega_deform_zoom}

\caption{
{\bf The band profile of the matrix.}
(a)~ 
The algebraic average and median along the diagonals 
of the $X_{nm}$ matrix versus ${\omega\equiv (E_n{-}E_m)}$. 
The vertical axis is normalized with respect to $C_{\infty}$, 
while the horizontal axis is $\omega/v_{\tbox{E}}$. 
The classical power spectrum is presented to demonstrate 
the applicability of the semiclassical relation \Eq{e2}.
The red line is the analytical expression 
that applies to zero deformation.
The quantum analysis is for~${R=8}$ with ${100<E<4000}$ (EW1), and with ${10000<E<14000}$ (EW2). 
The dotted vertical line is the frequency $1/t_{\tbox{L}}$ and the dashed one is $1/t_{\tbox{R}}$.
(b)~ 
Zoom of the $\omega\ll 1/t_{\tbox{L}}$ region.
For sake of comparison we display results also for~${R=2}$.  
The vertical lines indicate the mean level spacing.
The dashed red curves are a refined version of Eq.(\ref{e5}).
}

\mylabel{fig:F:alg:vs:omega:deform}

\mylabel{fig:F:alg:vs:omega:deform:zoom}

\end{figure}

\vspace{2mm}

{\bf \rmrk{The absorption coefficient.--} } 
\rmrk{Having in mind cold atoms in an optical trap},  
we regard the wall vibrations, say of the ``piston", 
as low frequency noisy driving. 
The power spectrum of~$\dot{f}(t)$ 
is described by a \rmrk{spectral function}
\be{0} 
\tilde{S}(\omega) \ \ = \ \ \varepsilon^2 
\, \frac{1}{2\omega_c}\exp\left(-\frac{|\omega|}{\omega_c}\right)
\ee  
As is common in the mesoscopic context we assume its spectral  
support to be $\omega_c \lesssim 1/t_{\tbox{R}}$, 
but larger compared with the mean level spacing.  
\rmrk{Accordingly, in the numerics} it is natural 
to take $\omega_c$ as matching the first minimum 
in the bandprofile of \Fig{fig:F:alg:vs:omega:deform}.

Following \mycite{kbw} we assume that there are FGR transitions between levels,
whose rate is proportional to $|F_{nm}|^2 \tilde{S}(E_n{-}E_m)$.
As a result the system absorbs energy in rate $G\varepsilon^2$ 
analogous to Joule heating. \rmrk{We define} 
\be{100}
G_0 \ \ = \ \ \frac{1}{2T}C_{\infty} 
\ \ \equiv \ \  
\frac{1}{2T} 
\left[\frac{8}{3\pi} \frac{\mass^2 v_{\tbox{E}}^3}{L_x}\right] 
\ee
This is the classical hard chaos result for the absorption coefficient, 
which is obtained, e.g. using a kinetic picture,  
if one neglects correlations between successive collisions. 
This is a straightforward adaptation of the well known ``Wall formula" of nuclear physics, 
which is analogous to the ``Drude formula" in condensed matter physics. 

\vspace{2mm}

{\bf \rmrk{Objective.--} } 
Our objective is to calculate the actual absorption coefficient~$G$, 
i.e. to go beyond the ``Wall formula" prediction, taking into 
account the implications of having ${t_{\tbox{R}} \gg t_{\tbox{L}}}$, 
which is the case for small deformation (${u\ll1}$). 
The calculation of the actual absorption coefficient~$G$ 
will be done below either within the framework of LRT 
using the Kubo formula (getting $G_{\tbox{LRT}}$), 
or within the framework of semi-linear response theory (SLRT) \mycite{kbr,slr,kbw}  
using a resistor-network calculation (getting $G_{\tbox{SLRT}}$). 
The correlations between collisions lead to 
an LRT result that we would like to write 
as ${G_{\tbox{LRT}}=g_c G_0}$.
Similarly it is convenient to write the outcome 
of the SLRT analysis as follows:
\be{0}
G_{\tbox{SLRT}} \ \ = \ \ g_s \, G_{\tbox{LRT}} \ \ = \ \ g_s \, g_c \, G_0 \ \ = \ \ g \, G_0 
\ee
If QCC considerations apply, then ${g_s\sim 1}$ 
with small $\hbar$ dependent corrections. 
The LRT and SLRT numerical results for $g$ are displayed 
in \Fig{fig:cqc:g:slrt:vs:E}, 
and the details are presented in what follows. 

\vspace{2mm}

{\bf Conflicting expectations.-- } 
\rmrk{Both in LRT and in SLRT the result for~$G$  
depends on the ``average" over the near diagonal 
elements of $|F_{nm}|^2$, i.e. those 
that are in the strip ${|E_n-E_m|\lesssim \omega_c}$.} 
The difference between LRT and SLRT is how this 
``average" is defined: as a simple algebraic average, 
or via a resistor network calculation. 
For a small deformation, first order perturbation 
theory (FOPT) implies that these couplings are $\propto u^2$.
But as $u$ becomes larger the common expectation, 
based on Wigner theory, is to have Lorentzian mixing of levels, 
leading to $\propto 1/u^2$ smearing. 
In the formally equivalent problem of a conductance 
calculation this implies ${G\propto 1/u^2}$, 
where $u$ represents the strength of the disordered potential 
(instead of using the FGR or Wigner picture one 
can use the equivalent Drude picture where 
the Born mean free path is ${\propto 1/u^2}$).    
On the other hand the semiclassical 
expectation, based on kinetic consideration, 
is to have, because of the bouncing, 
enhanced energy absorption ${\propto 1/u}$.  
Loosely speaking the latter expectation   
follows from the observation that a sequence 
of $1/u$~correlated collisions with the piston 
is like a single big collision.  
The purpose of the following paragraphs is to resolve 
this confusion by adopting a generalized random matrix theory (RMT) perspective.

\begin{figure}
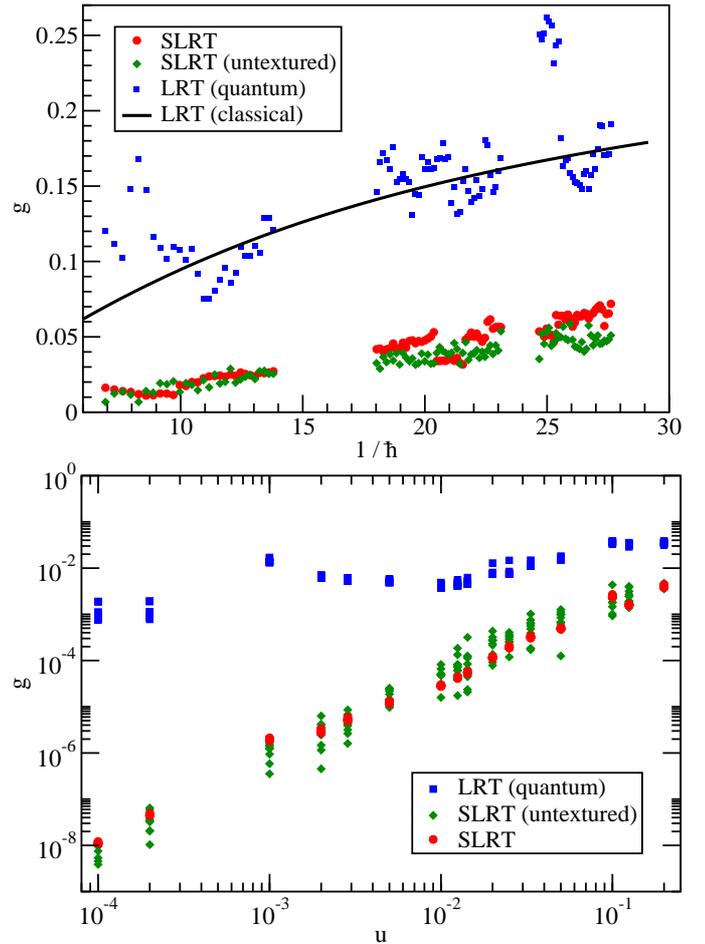

\centering
\includegraphics[clip,width=\hsize]{cqb_g_vs_h}

\includegraphics[clip,width=\hsize]{cqc_g_vs_u}

\caption{
{\bf SLRT vs LRT.}
The scaled absorption coefficient $g_c$ (LRT) and $g=g_sg_c$ (SLRT) 
versus the dimensionless $1/\hbar$ (upper panel),
and versus the dimensionless deformation parameter ${u=L/R}$ (lower panel).
Note that ${g=1}$ is the prediction of the ``Wall formula", 
while the line is based on the {\em classical} analysis.
In the upper panel the analysis has been done for the billiard of \Fig{fig:cqc:billiard}. 
The calculation of each point has been carried out on a $100 \times 100$ 
sub-matrix of $\bm{X}$  centered around the $\hbar$ implied energy~$E$. 
The ``untextured'' data points are calculated for an artificial  
random matrices with the same bandprofile and sparsity (but no texture).  
The complementary lower panel is oriented to show the small~$u$ 
dependence. The analysis is based on a truncated matrix 
representation of ${\mathcal{H}_0+U}$,  
within an energy window that corresponds to ${1/\hbar\sim9}$. 
Due to the truncation there is some quantitative inaccuracy 
with regard to the larger~$g$ values.   
}

\mylabel{fig:cqc:g:slrt:vs:E}
\end{figure}

\vspace{2mm}

{\bf RMT modeling.-- } 
So called ``quantum chaos" is the study of quantized
chaotic systems. Assuming that the classical dynamics
is fully chaotic, as in the case of a billiard 
with convex walls (\Fig{fig:cqc:billiard}), one expects the Hamiltonian 
to be like a random matrix with elements that have a Gaussian
distribution. This is of course a sloppy statement,
since any Hamiltonian is diagonal in some basis. 
The more precise statement is following~\mycite{mario1}:
Assume that $\mathcal{H}$ generates chaotic dynamics, 
and consider an observable $F$ that has some classical  
correlation function $C(t)$, with some correlation time $t_{\tbox{R}}$.
Then the matrix representation $F_{nm}$ in the basis
of $\mathcal{H}$ looks like a random banded matrix. 
The bandwidth is $\hbar/t_{\tbox{R}}$. 
If $t_{\tbox{R}}$ is small, such that the bandwidth 
is large compared with the energy window of interest, 
then the matrix looks like it is taken from a Gaussian ensemble.

What emerges in our numerical example, 
we would like to call ``weak quantum chaos" (WQC) circumstances, 
for which the traditional RMT modeling does not apply. 
Namely, in such circumstances it is not enough to characterize~$F_{nm}$ 
by its semiclassically-determined {\em bandprofile}. 
Rather one should further characterize~$F_{nm}$ 
by its quantum-mechanically-determined {\em sparsity}~\mycite{sparse1} 
and by its {\em texture}.    

\vspace{2mm}

{\bf Bandprofile.-- } 
Define a matrix $\bm{X}$ whose elements are~${X_{nm}=|F_{nm}|^2}$. 
The bandprofile $\bar{C}_a(r)$ is obtained by averaging 
the elements $X_{nm}$ along the diagonals ${n{-}m=r}$,  
within the energy window of interest.
In the same way we also define a {\em median} based bandprofile $\bar{C}_s(r)$.
See \Fig{fig:F:alg:vs:omega:deform} for numerical results. 
The mean level spacing is 
\be{0}
\Delta_0 \ \ = \ \ 2\pi/(\mass L_xL_y)
\ee
Given that $\Delta_0$ is small compared with the energy range of interest, 
it is well known~\mycite{mario1} that 
\be{2}
\bar{C}_a(n-m) \ \ = \ \ \left(\frac{2\pi}{\Delta_0}\right)^{-1} \ \tilde{C}(E_n-E_m) 
\ee
where $\tilde{C}(\omega)$ is the classical power spectrum, 
that can be obtained via the Fourier transform (FT) of the 
classical auto-correlation function $\langle F(0)F(t)\rangle$. 
\rmrk{In the numerical analysis} $F(t)$ corresponds 
to a very long ergodic trajectory. It consists of impulses, namely 
\be{6}
F(t) \ \ = \ \  \sum_j 2\mass v_{\tbox{E}} \ \cos(\theta_j) \ \delta(t-t_j) 
\ee
where $\theta_j$ is the collision angle with the piston at time~$t_j$. 
By the Wiener-Khinchin theorem $\tilde{C}(\omega)\propto|F_{\omega}|^2$, 
where ${F_{\omega}=\mbox{FT}[F(t)]}$. 
For technical details see~\mycite{dil}.  
The result of the calculation is displayed in \Fig{fig:F:alg:vs:omega:deform} 
(black continuous line). Comparing with the quantum one observes 
that the applicability of \Eq{e2} to the analysis of our billiard system 
is confirmed down to very small frequencies.

Analytical results for $\tilde{C}(\omega)$ can be obtained.
For large frequencies the power spectrum becomes flat and reaches 
the constant value~\mycite{kbw}
\be{3}
C(\omega \gg 1/t_{\tbox{L}}) 
\ = \ 
\frac{8}{3\pi} \frac{\mass^2 v_{\tbox{E}}^3}{L_x} 
\ \equiv \ 
C_{\infty}
\ee
For intermediate frequencies the effect of the deformation is 
mainly to ergodize the collision angle and one can obtain
analytical expression (represented in \Fig{fig:F:alg:vs:omega:deform} by dashed red line). 
For small frequencies the effect of the deformation 
is  less trivial and we find that 
the power spectrum is logarithmically divergent:
\be{5}
\tilde C(\omega \ll 1/t_{\tbox{L}}) \approx 
\mass^2  v_{\tbox{E}}^3 \ 
\frac{R}{2 L_x^2} \ 
\ln \frac{2}{\omega t_{\tbox{R}}}
\ee
The divergence comes because there are vertically bouncing 
trajectories with very long horizontal bouncing period, 
as in the related analysis of~\mycite{bouncing1}. 
\rmrk{Disregarding the logarithmic term one observe that 
compared with $C_{\infty}$ the bouncing leads to 
enhancement by factor~$1/u$, which is the ratio ${t_{\tbox{R}}/t_{\tbox{L}}}$. }

\begin{figure}
\centering
\includegraphics[clip,width=\hsize]{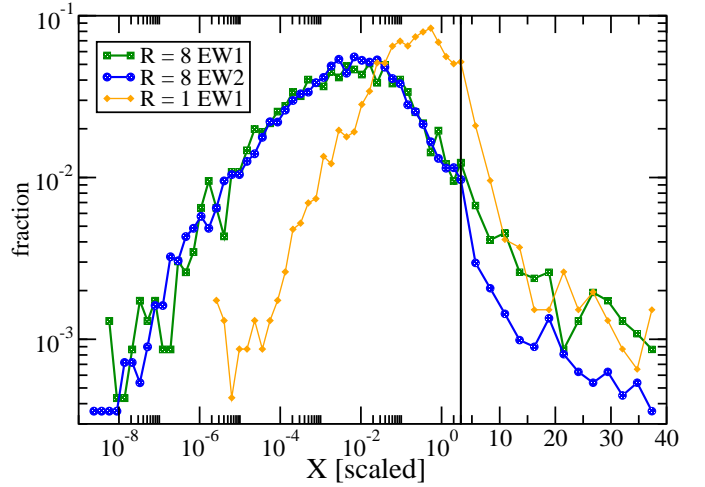}

\caption{
{\bf The size distribution of the matrix.}
Histogram of the values of $X_{nm}$  
for the central band of the EW1 and EW2 matrices 
as defined in \Fig{fig:F:alg:vs:omega:deform}.
For sake of comparison we display results also for~${R=1}$.  
}

\mylabel{fig:cqc:cumhist:F:nm}
\end{figure}

\vspace{2mm}

{\bf Sparsity and Texture.-- }
For strongly chaotic systems the elements within the band 
have approximately a Gaussian distribution. But in the WQC regime 
the matrix becomes {\em sparse} and {\em textured} 
as demonstrated in \Fig{fig:cqc:F:nm2:image}.
Loosely speaking, sparsity means that only 
a small fraction (${s\ll1}$) of elements are large\footnote{
A precise definition of the sparsity~$s$ can be found 
in Section~III of~\mycite{kbd}, 
\rmrk{but it is not of much physical interest for us}.
Rather we characterize the sparsity by the 
resistor-network measure~$g_s$ as defined below, 
which has direct relation to the response analysis.}, 
while the texture refers to their non-random arrangement.
In the WQC regime the size distribution of 
the in-band elements becomes log-wide (approximately log-normal) 
as seen in \Fig{fig:cqc:cumhist:F:nm}. 
This is reflected by having 
\be{0}
\bar{C}_s(r) \ \ \ll \ \ \bar{C}_a(r)
\ee
as seen in \Fig{fig:F:alg:vs:omega:deform}.

The sparsity and the texture of $\bm{X}$ are important for the analysis 
of the energy absorption rate~\mycite{kbw} 
as implied by SLRT~\mycite{kbr,slr}. Accordingly, 
we suggest to characterize the sparsity by a resistor network measure
\be{11} 
g_s \ \ = \ \ g_s[\bm{X}] \ \ \equiv \ \ {\langle\langle \bm{X} \rangle\rangle_s}/{\langle\langle \bm{X} \rangle\rangle_a}
\ee
Here $\langle\langle \bm{X} \rangle\rangle_a$ is the algebraic average 
over the in-band elements of the matrix, while $\langle\langle \bm{X} \rangle\rangle_s$
is the corresponding resistor network ``average" that takes their 
connectivity into account.
\rmrk{The recipe of the resistor network calculation is detailed in the 
next paragraph (can be skipped in first reading).}
For a strictly uniform matrix~${g_s=s=1}$, 
for a Gaussian matrix ${s=1/3}$ and ${g_s\sim1}$, 
while for sparse matrix ${s,g_s\ll1}$. 
 
The resistor network quantity $\langle\langle \bm{X} \rangle\rangle_s$  
can be regarded as a smart average over the elements of~$\bm{X}$,  
that takes their connectivity into account. For the purpose of its calculation  
we associate with $\bm{X}$ a matrix $\bm{\mathsf{g}}$ whose elements are  
\be{10}
\mathsf{g}_{nm} = 2\delta_0(n{-}m) \, \frac{X_{nm}}{(n-m)^2}
\ee
where $\sum_r \delta_0(r)=1$ is a weight function,  
whose width should be quantum mechanically large 
(i.e.~$\gg1$) but semiclassically small (i.e.~$\lesssim$~the~bandwidth).
If we take this weight function to be the normalized 
version of $\tilde{S}(\omega)$, then $\mathsf{g}_{nm}$ 
can be interpreted as the (normalized) Fermi-golden rule transition 
rates that would be induced by a low-frequency driving. 
Optionally we can regard these $\mathsf{g}_{nm}$ as representing 
connectors in a resistor network. The inverse 
resistivity of the strip can be calculated using the 
standard procedure, as in electrical engineering, 
and the result we call $\langle\langle \bm{X} \rangle\rangle_s$. 
It is useful to notice that if all the elements 
of $\bm{X}$ are identical, then $\langle\langle \bm{X} \rangle\rangle_s$ equals the same number.
More generally $\langle\langle \bm{X} \rangle\rangle_s$ 
is smaller than the conventional algebraic average $\langle\langle \bm{X} \rangle\rangle_a$ 
(calculated with the same weight function). 
In the RMT context a realistic estimate for $\langle\langle \bm{X} \rangle\rangle_s$ can 
be obtained using a generalized variable-range-hopping procedure~\mycite{kbd}.

\vspace{2mm}

{\bf The WQC regime.-- } 
With the classical $t_{\tbox{L}}$ and $t_{\tbox{R}}$, 
we can associate the energies 
\be{0}
\Delta_{\tbox{L}} &=& 2\pi/t_{\tbox{L}} \\ 
\Delta_{\tbox{R}} &=& 2\pi/t_{\tbox{R}}
\ee
Conversely, with the mean levels spacing we can associate 
the Heisenberg time $t_{\tbox{H}}=2\pi/\Delta_0$.
Note that $t_{\tbox{H}}=(1/\hbar)^{d{-}1}t_{\tbox{L}}$ where $d{=}2$.
It is also possible to define the Ehernfest time $t_{\tbox{E}}=[\log(1/\hbar)] t_{\tbox{R}}$, 
which is the time required for the instability to show up in the 
quantum dynamics. 
The traditional condition for ``quantum chaos"  
is ${t_{\tbox{E}} \ll t_{\tbox{H}}}$, 
but if we neglect the log factor it is simply ${t_{\tbox{R}} \ll t_{\tbox{H}}}$. 
This can be rewritten as ${\Delta_{\tbox{R}} \gg \Delta_0}$, 
which we call the frequency domain version of the quantum chaos condition. 
Optionally one may write a {\em parametric version} 
of the quantum chaos condition, namely ${u \gg u_b}$, where 
\be{0}
u_b = \hbar  \ \ \ \ \ \ \ \mbox{[de-Broglie deformation]}
\ee
The frequency domain version implies that 
it should be possible to resolve the zero frequency 
peak of $\tilde{C}(\omega)$ 
as in \Fig{fig:F:alg:vs:omega:deform}, 
while the parametric version means 
that a de-Broglie wavelength deformation of the boundary  
is required to achieve ``Quantum chaos".

We observe in the upper panel of \Fig{fig:cqc:g:slrt:vs:E}  
that ${g_s}$ is significantly smaller than unity, 
even for very small values of~$\hbar$ 
for which ${u > u_b}$ is definitely satisfied.  
For completeness we show in the lower plot additional 
data points in the regime ${u < u_b}$ where this breakdown 
of QCC is not a big surprise. 
We conclude that QCC for ${u > u_b}$ is restricted to $\tilde{C}(\omega)$, 
and does not imply {\em Hard} quantum chaos (HQC), 
but only WQC. In the WQC regime ${\bar{C}_s(r)\ll \bar{C}_a(r)}$
and consequently ${g_s \ll 1}$, indicating sparsity.

The emergence of WQC instead of HQC can be explained as follows.  
If a wall of a billiard is deformed, the levels are mixed.
FOPT is valid provided  ${|U_{nm}| < \Delta_0}$. 
This condition determines a parametric scale~$u_c$. 
If the unperturbed billiard were chaotic, the variation 
required for level mixing would be~\mycite{prm} 
$u_c \approx \lambda_{\tbox{E}}/(k_{\tbox{E}}L)^{1/2} = \hbar^{3/2}$. 
This expression assumes that the eigenstates look like random waves.  
In the Wigner regime (${u_c<u<u_b}$) there is 
a Lorentzian mixing of the levels and accordingly,  
the number of mixed levels is $\sim(u/u_c)^2$. 
But our unperturbed (rectangular) billiard is not chaotic, 
the unperturbed levels of the non-deformed billiards
are not like random waves.  
Therefore, the mixing of the levels is {\em non-uniform}.

By inspection of the $U_{n_xn_y,m_xm_y}$ matrix elements 
one observes that the dominant matrix elements that are responsible 
for the mixing are those with large~${n_x}$ but small ${|n_y{-}m_y|}$. 
Accordingly, within the energy shell ${E_{n_xn_y} \sim E}$, 
the levels that are mixed first are those with maximal ${n_x}$, 
while those those with minimal ${n_x}$ are mixed last. 
The mixing threshold for the former is 
\be{0}
u_c \ \ \approx \ \ \lambda_{\tbox{E}}/(k_{\tbox{E}}L) \ \ = \ \ \hbar^{2}
\ee
while for the latter one finds  $u_c^{\infty} \sim \hbar^{0}$, 
which is much larger than $u_b=\hbar^{1}$. 
Straightforward analysis of this mixing (extending that of~\mycite{kbw}) 
leads to the result 
\be{0}
g \ \ \approx \ \ u^2/\hbar 
\ee
This is merely the ratio of the median value to the mean, 
and the proportionality to~$u^2$ is the remnant of FOPT.
This simple dependence is confirmed by the numerics of \Fig{fig:cqc:g:slrt:vs:E}. 
We note that the RMT perspective of~\cite{kbd} implies 
that in general this median based estimate should be 
corrected. Roughly the prescription is 
\be{0}
g \ \ \mapsto \ \ \max\{1,g \exp\left[\sqrt{-\ln b \ln g} \right]\}
\ee
where ${b=\omega_c/\Delta_0}$ is the dimensionless bandwidth.

\rmrk{In the numerics $g$ is calculated for a bandwidth 
matching spectral width,} i.e. the spectral support 
of $\tilde{S}(\omega)$ is assumed to be $\sim \Delta_{\tbox{R}}$, 
implying ${g_c\sim \mathcal{O}(1)}$ and ${g \sim g_s}$. 
In the quantum mechanical LRT calculation which is presented 
in \Fig{fig:cqc:g:slrt:vs:E} by black line $g_c$ depends 
on $\hbar$, because $\Delta_0$ provides a lower cutoff on 
the logarithmically divergent $\tilde{C}(\omega)$.

If the spectral support of the driving were $\ll \Delta_{\tbox{R}}$, 
the classical correlation factor would be ${g_c \sim 1/u}$, 
and consequently $g_s\sim u^3/\hbar$.
Still, the bandwidth is the significant scale in  
the ``quantum chaos" perspective, and therefore 
the parametric scale that signifies the WQC-HQC crossover is  
\be{325}
u_s \ \ = \ \ \hbar^{1/2}
\ee
which is larger than ${u_b=\hbar}$.      
Accordingly, the WQC regime extends well beyond the 
traditional boundary of the Wigner regime, 
and in any case it is well beyond the FOPT border~$u_c$.

\vspace{2mm}

{\bf \rmrk{Discussion.--} } 
In a broader perspective the term ``weak quantum chaos" is possibly 
appropriate also to system with zero Lyapunov exponent ($t_{\tbox{R}}{=}\infty$), 
e.g. the triangular billiard~\mycite{triang}, 
and pseudointegrable billiards~\mycite{spectral}, 
and to systems with a classical mixed phase space. 
\rmrk{But in the present study} we wanted to consider 
a globally chaotic system, under semiclassical circumstances such 
that $\Delta_{\tbox{R}}$ is quantum mechanically resolved and QCC is naively expected. 
In this context there are of course other interesting aspects, 
such as bouncing related corrections to Weyl's law~\mycite{Backer}, 
and non-universal spectral statistics issues (see below), 
while our interest was with regard to the semi-linear response 
characteristics of the system. 

\rmrk{The spectral statistics in the WQC regime} has been 
studied in~\mycite{wqc1} concerning nearly circular stadium billiard, 
and in~\mycite{wqc2} concerning circular billiards with a rough boundary. 
Let us remind very briefly how the WQC border is determined 
in this context. It is convenient to describe the dynamics 
using a Poincare map, which relates the angle $\theta_{\tau}$  
of successive collisions (${\tau=1,2,3,\cdots}$) with the piston.
One observes that due to the accumulated effect of 
collisions with the deformed boundary, there is a slow diffusion  
of the angle with coefficient ${D_{\theta}\sim u^2}$.
Accordingly the classical ergodic time is ${\tau_{r} \sim 1/D_{\theta}}$, 
and the quantum breaktime due to a dynamical localization 
effect is ${\tau_{h} \sim D_{\theta}/\hbar^2}$. 
The border of the WQC regime is defined by the 
condition ${\tau_{h}<\tau_{r}}$ leading to~\Eq{e325}.
However we would not like to over-emphasize this consistency 
because it is not a-priori clear that spectral-statistics
and sparsity related characteristics always coincide.

\vspace{2mm}

{\bf Practical implications.-- } 
\rmrk{Coming back to the ``conflicting expectations" issue,} 
with regard to the value of the absorption coefficient and its dependence 
on the deformation~$u$, we now can see how they reconcile.   
First of all it should be clear that if there were no classical 
correlations between bounces, then $\tilde{C}(\omega)$ would be flat, 
equals to the value $C_{\infty}$ of \Eq{e3}, 
leading to the wall formula \Eq{e100} for~$G$.
The effect of bouncing is to enhance $\tilde C(\omega \ll 1/t_{\tbox{L}})$ 
as implied by \Eq{e5}. Depending on whether the spectral 
support of the driving is $\omega_c\ll \Delta_{\tbox{R}}$
or $\omega_c\sim \Delta_{\tbox{R}}$ we observe or do not observe a $1/u$ enhancement. 
This holds classically and also in the quantum LRT calculation  
(provided ${\omega_c > \Delta_0}$) due to QCC.

However, the SLRT calculation, unlike the LRT calculation, cares 
about the {\em median} and not about the {\em mean}. 
Therefore, for a weakly chaotic system, 
it give a much smaller result for~$G$. If the mixing of the levels 
were uniform we would expect a crossover from $g_s\propto u^2$ (FOPT) 
to  $g_s \propto 1/u^2$ (Wigner), as in the theory of disordered conductors. 
But the mixing of levels in a weakly {\em chaotic} system, 
unlike in a weakly {\em disordered} system, is not uniform, 
and therefore the $g_s \propto u^2$ persists within 
a very large range $u_c<u<u_s$, to which we refer as the WQC regime.

\vspace{2mm}

{\bf \rmrk{Experimental feasibility}.-- } 
Having a better understanding of the WQC regime we are 
now able to revise the suggested experiment in~\cite{kbw}.  
Let us consider ${}^{85}Rb$ atoms that are laser cooled 
to low temperature ${T \approx 0.1\mu K}$, 
such that the de-Broglie wavelength is ${\lambda_{\tbox{E}} = 1 \,\mu m}$. 
The atoms are trapped in an optical billiard of linear 
size of  ${L=10 \,\mu m}$, and accordingly    
the dimensionless Planck constant is $\hbar=0.1$. 
This leads to $\Delta_{\tbox{L}}/\Delta_0=30$. 
Note that ${\Delta_{\tbox{L}}=220 \,\mbox{Hz}}$, 
and ${\Delta_0 =7.5 \,\mbox{Hz}}$. 

Assuming $10\%$ deformation the dimensionless bandwidth 
can be tuned as $b \equiv (\Delta_{\tbox{R}}/\Delta_0) \sim 10$.   
By modulating the laser intensity, 
one of the billiard walls can be noisily vibrated.
We assume that the driving is band-matching, 
i.e. $\omega_c\sim \Delta_{\tbox{R}}$.
These are roughly the same parameters as in our numerical analysis.
The prediction for the SLRT suppression factor is ${g_s\sim 0.1}$.

In order to witness the SLRT anomaly the RMS amplitude of the 
vibrations ($\varepsilon$) should be large enough, 
as to have a measurable heating effect.
Assuming that it is possible to hold the atoms 
for a duration of $\sim 1000$ bounces the condition 
can be written in a dimensionless form 
as ${G_0 \varepsilon^2 / (T \Delta_{\tbox{L}}) > 10^{-3}}$, 
or roughly as ${(\varepsilon/L)^2 > 10^{-3}}$.

On the other hand $\varepsilon$ should be small enough, 
such that the FGR condition is not violated. 
It is straightforward to show that the FGR condition  
can be written in a dimensionless form 
as ${TG_0 \varepsilon^2/\Delta_0^3 < b^3}$, 
or roughly as ${(\varepsilon/L) < (1/b)}$.
Accordingly there is a range where both conditions 
are satisfied, and there the SLRT anomaly should be observed, 
provided environmental relaxation effects can be neglected.

\vspace{2mm}

{\bf Comments.-- } 
It is important to realize that we are studying in this work 
a driven chaotic system, and not a driven integrable system. 
Remarkable examples for driven integrable systems are the 
kicked rotator \mycite{kr} and the vibrating elliptical billiard \mycite{eb}.
In the absence of driving such systems are integrable, 
while in the presence of driving a {\em mixed phase space} emerges.      
This is not what we call here {\em weak chaos}. 

\rmrk{The low frequency driving} that we assume is stochastic, 
rather than periodic. This looks to us realistic, 
reflecting the physics of cold atoms that are trapped in optical billiards with vibrating walls. 
It is also theoretically convenient, because we can use the FGR picture. 
If one is interested in periodic driving of strictly isolated system, 
then there are additional important questions  
with regard to dynamical localization \mycite{lc}, 
that can be handled e.g. within the framework of the Floquet theory approach. 

\clearpage

{\bf Summary.-- } 
The discovery of ``anomalies", i.e. major deviations from QCC  
in circumstances where QCC is expected by common wisdom, 
is a major challenge in quantum-mechanics studies.  
For example: Anderson's Localization (wavefucntions were commonly expected to be extended); 
Heller's scars (wavefucntions were commonly expected to look like random waves). 
Here we highlighted an anomaly in the theory of response: 
the rate of heating is unexpectedly suppressed for a quantized chaotic system. 

Our analysis has been based on SLRT. This theory applies 
to circumstances in which the environmental relaxation 
is weak compared with the $f(t)$-induced transitions. 
In such circumstances the connectivity of the transitions from level to level 
is important, and the LRT result should be multiplied by~$g_s$.      

We have highlighted that there is a {\em distinct WQC regime}, 
where semiclassics and Wigner-type mixing co-exist. 
This is the regime where an LRT to SLRT crossover is expected 
as the intensity of the driving is increased.



{\bf Acknowledgements.-- }
\rmrk{We thank Nir Davidson (Weizmann) for a crucial 
discussion regarding the experimental details.}  
This research has been supported by the US-Israel Binational Science Foundation (BSF). 


\clearpage
\end{document}